\documentstyle[psfig]{mn}

\newif\ifAMStwofonts

\ifoldfss
  \ifCUPmtlplainloaded \else
    \NewTextAlphabet{textbfit} {cmbxti10} {}
    \NewTextAlphabet{textbfss} {cmssbx10} {}
    \NewMathAlphabet{mathbfit} {cmbxti10} {} 
    \NewMathAlphabet{mathbfss} {cmssbx10} {} 
  \fi
  \ifAMStwofonts
    \ifCUPmtlplainloaded \else
      \NewSymbolFont{upmath} {eurm10}
      \NewSymbolFont{AMSa} {msam10}
      \NewMathSymbol{\upi}     {0}{upmath}{19}
      \NewMathSymbol{\umu}     {0}{upmath}{16}
      \NewMathSymbol{\upartial}{0}{upmath}{40}
      \NewMathSymbol{\leqslant}{3}{AMSa}{36}
      \NewMathSymbol{\geqslant}{3}{AMSa}{3E}

       \let\ge=\geqslant
    \fi
  \fi
\fi 

\ifnfssone
  \newmathalphabet{\mathit}
  \addtoversion{normal}{\mathit}{cmr}{m}{it}
  \addtoversion{bold}{\mathit}{cmr}{bx}{it}
  \newmathalphabet{\mathbfit} 
  \addtoversion{normal}{\mathbfit}{cmr}{bx}{it}
  \addtoversion{bold}{\mathbfit}{cmr}{bx}{it}
  \newmathalphabet{\mathbfss} 
  \addtoversion{normal}{\mathbfss}{cmss}{bx}{n}
  \addtoversion{bold}{\mathbfss}{cmss}{bx}{n}
  \ifAMStwofonts
    \ifCUPmtlplainloaded \else
      \UseAMStwoboldmath
      \makeatletter
      \new@mathgroup\upmath@group
      \define@mathgroup\mv@normal\upmath@group{eur}{m}{n}
      \define@mathgroup\mv@bold\upmath@group{eur}{b}{n}
      \edef\UPM{\hexnumber\upmath@group}
      \new@mathgroup\amsa@group
      \define@mathgroup\mv@normal\amsa@group{msa}{m}{n}
      \define@mathgroup\mv@bold\amsa@group{msa}{m}{n}
      \edef\AMSa{\hexnumber\amsa@group}
      \makeatother
      \mathchardef\upi="0\UPM19
      \mathchardef\umu="0\UPM16
      \mathchardef\upartial="0\UPM40
      \mathchardef\leqslant="3\AMSa36
      \mathchardef\geqslant="3\AMSa3E

       \let\ge=\geqslant
    \fi
  \fi
\fi 

\ifnfsstwo
  \DeclareMathAlphabet{\mathbfit}{OT1}{cmr}{bx}{it}
  \SetMathAlphabet\mathbfit{bold}{OT1}{cmr}{bx}{it}
  \DeclareMathAlphabet{\mathbfss}{OT1}{cmss}{bx}{n}
  \SetMathAlphabet\mathbfss{bold}{OT1}{cmss}{bx}{n}
  \ifAMStwofonts
    \ifCUPmtlplainloaded \else
      \DeclareSymbolFont{UPM}{U}{eur}{m}{n}
      \SetSymbolFont{UPM}{bold}{U}{eur}{b}{n}
      \DeclareSymbolFont{AMSa}{U}{msa}{m}{n}
      \DeclareMathSymbol{\upi}{0}{UPM}{"19}
      \DeclareMathSymbol{\umu}{0}{UPM}{"16}
      \DeclareMathSymbol{\upartial}{0}{UPM}{"40}
      \DeclareMathSymbol{\leqslant}{3}{AMSa}{"36}
      \DeclareMathSymbol{\geqslant}{3}{AMSa}{"3E}

       \let\ge=\geqslant
    \fi
  \fi
\fi 

\ifCUPmtlplainloaded \else
  \ifAMStwofonts \else 
    \def\upi{\pi}
    \def\umu{\mu}
    \def\upartial{\partial}
  \fi
\fi

\title [CF~Octantis, 1964--1976]
{Archival light curves from the Bamberg Sky Patrol -- CF~Octantis, 1964--1976} 

\author[J.L. Innis, A.P. Borisova, D.W. Coates, and M.K. Tsvetkov]
{J.L. Innis$^{1}$, A.P Borisova$^{2}$, D.W. Coates$^{3}$ and M.K. 
Tsvetkov$^{2}$\\
$^{1}$ 43 Ash Drive, Kingston, Tasmania, 7050, Australia, j.innis@aip.org.au\\
$^{2}$ Institute of Astronomy, 72 Tsarigradsko Shosse Blvd., 1784, Sofia, 
Bulgaria, ana@skyarchive.org, 
milcho@skyarchive.org\\
$^{3}$ School of Physics and Materials Engineering, Building 27, Monash 
University, Victoria, 3800, Australia, denis.coates@spme.monash.edu.au\\}
\date{22 July 2004}

\pagerange{\pageref{firstpage} -- \pageref{lastpage}}
\pubyear{0000}
\begin{document}

\maketitle

\label{firstpage}

\begin{abstract}
 
We use the archive of the Bamberg Sky Patrol to obtain light curves of the 
active K subgiant CF~Octantis for the interval 1964-1976.  Digitised images of 
the field near CF~Oct were obtained with a flat--bed scanner. Aperture 
photometry was performed of photo--positives of these images. Using a 
transformation to second order in plate magnitude, and first order in {\it 
B}$-${\it V}, 
for 9 field stars for each plate, the {\it B} magnitudes of CF~Octantis were 
obtained for just over 350 plates. The estimated precision of an individual 
determination of the {\it B} magnitude of CF~Oct is 0.05~mag. Analysis of the 
resulting data reveals the known 20~d rotational variation of this star, and 
shows the evolution of the light curves from year to year.  We obtain light 
curves with good phase coverage for 1964 to 1969 inclusive, partial light 
curves for 1970 and 1976, and a few data points from 1971. The amplitude of 
variation ranges from $\sim$0.2 to $\sim$0.4 mag. There is evidence that the 
characteristic rotation period of the star in the 1960s was slightly less than 
that measured from photoelectric photometry in the 1980s.
\end{abstract}


\section{Introduction} 

\label{introduction}
Detailed studies of astrophysically interesting objects often require 
long--term datasets to fully understand the nature of any variations.  There 
is a growing awareness that the archives of astronomical photographic plates 
potentially contain an enormous quantity of information, extending, in some 
cases, over several decades or longer (e.g. Kroll et al, 1999).  The utility 
of photographic plate archives for the long--term study of active stars has 
been demonstrated by, for example, the work of Phillips \& Hartmann (1978), 
Hartmann et al. (1981), and Bondar (1995).  These works present data gleaned 
from plate archives showing the longer term (decadal) variation of a number 
of active stars, revealing strong evidence for starspot cycles in some 
objects.  The stars studied were mostly BY Dra objects (dKe or dMe flare 
stars), although one star, PZ Mon, has since been shown to be a K giant, and 
probably a member of the RS~CVn class (Saar, 1998).

In this work, we use the Southern Sky Patrol plate archive of the Dr. Remeis 
Sternwarte, Bamberg, Germany (hereafter referred to as the Bamberg 
Observatory) to investigate the behaviour of a selected star, using digitised 
plate images obtained with a fast scanner. Our target object is the active, 
probably single, K subgiant CF~Octantis.  The long--term brightness variations 
of the star can be followed, and seasonal light curves can be found.  The 
techniques we use should be applicable to a variety of variable stars of 
moderately large amplitude ($\ge$0.15~mag).

In the next section we briefly describe the Bamberg sky patrol programme, 
provide background information on our choice of target star for this study, 
and outline how the plates were scanned and {\it B} magnitudes obtained from 
the digital data.  Section~\ref{discussion} presents a discussion of our 
results, including estimates of the accuracy and precision of the photometry, 
and an assessment of the wider applicability of this approach.

This paper is mainly concerned with the techniques used to obtain photometric 
data from the archival plates, using CF~Oct as an example. A more 
detailed discussion of the scientific findings will be presented later.  An 
earlier paper (Innis et al., 2004) presented photometric measurements of 
CF~Oct from a small subset of the data presented here, but in that case 
the photometry was from digital camera images of the plates, obtained 
before the plate scanner was installed at Bamberg Observatory.  In the last 
section of this paper we will briefly compare the results from the two 
techniques.
 
\section{Data} 
\label{data}

\subsection{Bamberg Sky Patrol} 

The Southern Sky patrol was undertaken by the Bamberg Observatory 
from 1963 to 1976, specifically to identify and monitor variable stars. The 
early plates were taken at Boyden, South Africa, but from 1967 the 
programme moved to stations at Mt John, New Zealand, and San Miguel, 
Argentina.  A bank of six cameras on a common mount obtained wide--field 
plates (approximately 13$^{\circ}\times$13$^{\circ}$).  The patrol plates were exposed 
for 1~hour, and recorded stars to magnitude 14. For further information see 
Strohmeier \& Mauder (1969). The blue-sensitive emulsion (Agfa Gevaert 67 
A50) produced data that were interpreted as photographic magnitudes 
(m$_{pg}$).  Many variable stars were found and characterised (e.g. Schoeffel 
\& Koehler, 1965).  The plates have yielded a photometric precision near 
0.05~mag, as demonstrated by iris photometer measurements by various workers 
(e.g. Schoffel, 1964). 

The plate archive is maintained in good condition by the Bamberg Observatory. 
An on--line catalogue of the archive is available at the Sofia Wide Field 
Plate Database at http://www.skyarchive.org. In mid 2003 a fast scanner was 
installed at Bamberg, allowing digital copies of the plates to be obtained.  
This paper presents some of the first results of use of this scanner.

\subsection{CF~Octantis (HD~196818)}

CF~Octantis (HD~196818, V$\sim$~8, $\alpha\sim$20$^{\rm h}$~50$^{\rm m}$, $\delta\sim-$80$^\circ$). 
was discovered to be variable by the Bamberg southern sky patrol, and was 
designated BV~893 (Strohmeier, 1967). At that time the period and the nature 
of the variability were not determined.  CF~Oct is now known to be a K0 star 
with a large and variable photometric rotational modulation 
($\sim$0.1--0.35~mag in V) with a $\sim$20.1~d period.  Photometry of this 
star has been published by Innis et al. (1983), Lloyd Evans \& Koen (1987), 
Pollard et al. (1989) and Innis, Coates and Thompson (1997). Spectroscopic 
studies include those of Hearnshaw (1979), Collier (1982), and Innis et al. 
(1997), and show the star exhibits strong Ca II emission, and a filled--in 
H$\alpha$ line.  The star is probably single, as no significant radial 
velocity variations have been found (Innis et al, 1997). The star is a strong, 
flaring, microwave radio source (Slee et al., 1987a, b).  The spectral type is 
K0~IIIp from Houk \& Cowley (1975). Innis et al. (1997) however suggested the 
star is more likely a subgiant. The Hipparcos distance of 200~pc, together 
with the representative {\it V} maximum near 8, lead to an M$_{\rm{V}}$ around 
1.5, between that of a subgiant (M$_{\rm{V}}$=3) and that of a giant 
(M$_{\rm{V}}$=0; Allen, 1973).

The high southerly declination of CF~Oct meant the star was well covered by 
the Bamberg sky patrol. One camera on the Bamberg mount was centred on 
$-$77$^{\circ}$~declination. Plates obtained with this camera on nominal centres of 
19$^{\rm h}$, 20$^{\rm h}$, 21$^{\rm h}$, and 22$^{\rm h}$ of right ascension all included 
CF~Oct and the 
surrounding field.  Hence on any given night, this star may appear on up to 
four plates, depending on the season and weather conditions.  The data to be 
presented here come from approximately 350~plates from 1964 to 1976.

\subsection{Scanning the field of CF~Octantis}

In May 2003 a flatbed Epson Expression 1640XL scanner with a transparency unit 
became operational at the Bamberg observatory. It provides an opportunity for 
photographic plate and film digitization with a spatial resolution of 
16~microns and a maximal size of A3 format. The output data are stored in 
14--bit (16383 grey scale levels) FITS format. Dark frame subtraction, flat 
fielding and changing from photo-negative to photo-positive image are part of 
the automated scanning process. Typical header data are telescope type and 
WFPDB identifier, object of interest, and coordinates of plate centre.

\begin{figure*} 
\psfig{file=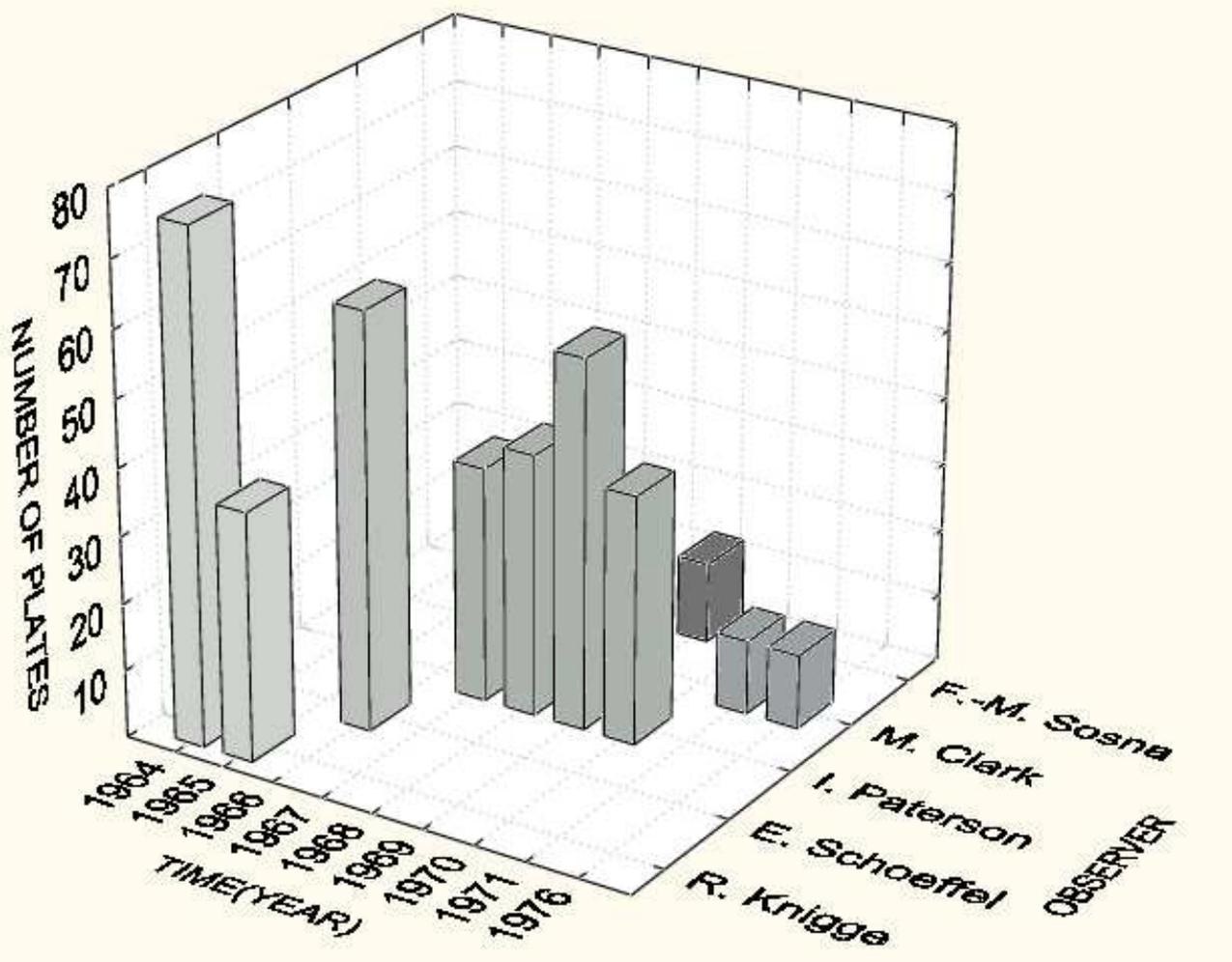,width=18cm}
\caption{Bar graph showing the scanned plates distributed by time and 
observer.}
\label{plates_sta}
\end{figure*}

For plates obtained with 10--cm astrographic cameras of the Bamberg 
observatory (at a scale of 338~arc~sec~mm$^{-1}$) the resulting resolution is 
5.25~arc~sec~pixel$^{-1}$.   A total of 375~fields centred on CF Oct were 
digitized in May and June 2003. Plates obtained in the period 1964-1966 are 
from Boyden Observatory (Republic of South Africa), 1967-1976 (with lack of 
plates in the period 1972-1975) are from Mount John University Observatory - 
Lake Tekapo (New Zealand), and 12 plates from 1969 are from San Miguel 
Observatory (Argentina). For the selected area of 2.75$^{\circ}$ $\times$ 
2.75$^{\circ}$ (4~cm $\times$ 4~cm) it took 1~minute for digitization. The 
file size is 8~Mb. The epochs of the scanned plates, and also the observers 
involved, are given in Figure~\ref{plates_sta}.

\subsection{Obtaining instrumental (plate) magnitudes with IRAF}

A number of workers have shown that aperture photometry of (photo--positive) 
images of digitised astronomical plates can yield satisfactory results if care 
is applied in the reduction procedures (e.g. Burkeholder, 1995; Vogt \& 
Kroll, 1999).  As is well known, the photometric response of the plate to 
incident light is significantly non--linear.  The reduction technique must 
take this into account.  It is apparent also that most (if not all) of the 
`modern' astronomical reduction packages have arisen with the advent of 
highly--linear digital detectors, and have not been designed with 
photographic plates in mind. Generally speaking these programs have enough 
flexibility to be adapted for use.  We have used the package IRAF for the 
analysis presented here, using the interactive task IMEXAMINE to obtain 
aperture photometry.

The star aperture radius and sky annulus radii were chosen to include the star 
image (from inspection of the radial profile plot of the brightest stars we 
measured), and to avoid contamination from nearby stars.  We typically used a 
20 pixel radius star aperture (hence an area $\sim$1250 pixels), with sky 
inner and outer annuli 25 and 30 pixels (area $\sim$860 pixels).

The brighter star images recorded on the scans showed flat--topped cores, indicative 
of saturation -- either in the emulsion or due to limitations of the scanner, 
with a very rapid fall to background.  The aperture photometry 
we perform is, in large part, therefore a measure of the dimensions of the 
saturated stellar image. Instrumental magnitudes are calculated in the conventional 
manner as 
\begin{equation}
M_{i}=-2.5 log(C) + K,
\end{equation}
where {\it C} are the `counts' recorded in the scanning process, and {\it K} is an 
arbitrary constant (the zero point) chosen for convenience so the M$_{i}$'s were comparable to their 
catalogued values. We used the same value of {\it K} for all plates.  The instrumental 
magnitudes must be carefully calibrated and transformed to a scale linear in {\it B} 
or m$_{pg}$ magnitude.  We outline below how this was accomplished.

The IMEXAMINE task has an iterative facility to determine the star's FWHM and 
adjust the aperture radius to suit. Possibly due to the non--gaussian stellar 
profiles, in a small number of instances the iterative procedure failed to 
converge.  Examination of other cases showed that, although convergence was 
obtained, the resulting flux sum, and hence derived magnitude, was clearly 
incorrect.  Consequently we turned off this iterative procedure, in effect 
fixing the aperture to the user--specified 20 pixels.  Inspection of the data 
showed signifcantly less scatter in the derived magnitudes with this procedure.

\subsection{Deriving the plate transformations}
\label{sec:fitting}
 
We measured nine (non--variable) field stars, of small (less 
than 2 degree) angular separation from CF~Oct, ranging in {\it B} mag from 
$\sim$8.7 to $\sim$10.7 to obtain for each plate a transformation equation 
from plate magnitude to {\it B} magnitude. We also derived transformation 
equations between plate magnitude and m$_{pg}$, the photographic magnitude. 
The transformations were equally good whether {\it B} or m$_{pg}$ was being 
used. We mostly refer to the B--magnitude results as we believe these have 
greater applicability.  The {\it B} magnitudes of the field stars have been 
determined by transforming the Tycho--2 BT and VT data (ESA, 1997), using 
interpolations of the tables presented by Bessell (2000).

It should be noted that this transformation was not only to provide a colour 
correction from the `natural' system of the emulsion and camera to standard {\it B} 
magnitudes, but also corrected for inherent non-linearities in 
the photographic technique.

\begin{figure}
\psfig{file=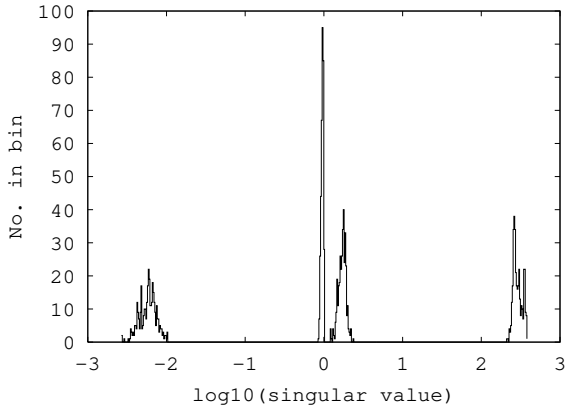,width=8cm,angle=270}
\caption{Histogram of the singular values found when applying the SVD method 
to the $\sim$350 plates.  The lowest cluster, centred near -2.2 in the log, 
were identified as arising from observational noise, hence the reciprocals of 
these values were set to zero when computing the SVD solution.}
\label{singular_values}
\end{figure}

For each plate a least--squares solution was found for the equation
\begin{equation}
\label{eq:with_colour}
 B = ~\beta_{1}M_{i} ~+~ \beta_{2}M_{i}^{2} ~+~ \beta_{3}(B$-$V) ~+~\beta_{4},
\end{equation}
where $\beta_{4}$ is an additive constant.  The least-squares solution was 
found using the Singular Value Decomposition (SVD) approach described by Press 
et al. (1992), and for the fits performed here two independent approaches were 
used, one being a FORTRAN implementation based on Press et al (1992), and the 
other using the OCTAVE data analysis package (by J.W. Eaton) which includes an 
SVD function. The results were effectively indistinguishable.   In applying 
the SVD method, it is necessary to set to zero the reciprocal of certain 
singular values in order to prevent observational errors (noise) dominating 
the solution. Our choice of which values to zero followed the approach 
recommended by Rucinski (1999), where the singular values are inspected, and 
an appropriate cut--off is determined.  For the configuration we employed, the 
application of the SVD procedure yielded 4 singular values for each plate. The 
distribution of the singular values is shown (on a semilog axis) in 
Figure~\ref{singular_values}.  There was a large measure of uniformity in the 
magnitude of the singular values found for each plate, as shown by the four 
clusters of data in this figure.  The lowest cluster, near $-$2.2 in the log 
($\sim$0.005) we identified as arising from observational noise.  These values 
would dominate the SVD solution (as the singular values are inverted in 
solving the equation), and hence their reciprocals were set to zero.  Doing 
so, the additive term, $\beta_4$, took on values near zero for all plate 
solutions, while not doing this resulted in the $\beta_4$ terms varying from 
$-$40 to $+$20.

The form of 
equation~\ref{eq:with_colour} was chosen after a 
number of trials, including using third order terms in M$_{i}$, and with and 
without an additive term.  The test in each case was how 
constant were the derived field--star magnitudes after applying the 
transformation equation to the plate magnitudes.  We found that a fit linear 
in M$_{i}$ and {\it B}$-${\it V} was significantly better 
than a second order fit in M${_i}$ with no colour term.  A five--parameter fit 
using terms of first and second order in both M$_{i}$ and {\it B}$-${\it V}, 
with an 
additive constant, was marginally better than that found with 
equation~\ref{eq:with_colour}, but the improvement 
was small.  For simplicity we have adopted equation~\ref{eq:with_colour} for 
use here. Our approach was developed independently of the earlier work of 
Kroll \& Neugebauer (1993), and Vogt \& Kroll (1999), but has reached 
essentially the same conclusions. However, Kroll \& Neugebauer (1993) and 
Vogt \& Kroll (1999) performed their fitting using a measure of the stellar 
flux (not the instrumental magnitude) derived from the stellar profile.

We defer further discussion of the nature of the colour correction to a later 
section.  However we note briefly that if no colour term is included the 
derived field star magnitudes can systematically vary by up to $\sim$0.5~mag 
over the data set.  These variations, also seen in the raw plate magnitudes, 
are removed almost entirely by inclusion of a colour term in the 
transformation equation. As shown later, the colour 
term, $\beta_{3}$, shows consistency within a season, but varies 
slowly from season to season.  This may relate to either ageing of the plates 
prior to exposure, or changes in the properties of the optical coatings on the 
camera lenses, or possibly even to atmospheric effects, or a combination of 
all of these.  We have also found a significant step in the values of this
term between 1966 and 1967, when the observing site changed from South Africa 
to New Zealand.  

\begin{figure} 
\psfig{file=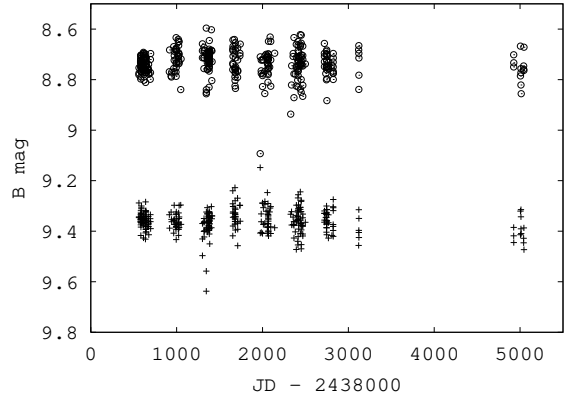,width=8cm}
\caption{ {\it B} magnitudes of HD~196520 (circles) and HD~192950 (upright 
crosses) 
vs JD, derived using plate transformations of the form of 
equation~\ref{eq:with_colour}.}
\label{colour}
\end{figure}

Once the $\beta$ terms have been found, the plate magnitudes of the field 
stars can be transformed to B~magnitudes. This is not an independent estimate 
of the field star magnitude, as of course the plate magnitudes form the input 
data used to derive the transformation. However, this allows the 
consistency of the technique to be examined. Figure~\ref{colour} shows 
the {\it B} magnitudes for HD~196520 and HD~192950 derived 
using equation~\ref{eq:with_colour}.  There is no evidence for variation above 
what appears to be observational scatter. We found similar results for all 
field stars.  There may be evidence for small systematic offsets between 
seasons at the $\sim$0.05~mag level, which presumably relates to the colour 
sensitivity of the plates, which is discussed below.  

\begin{figure}
\psfig{file=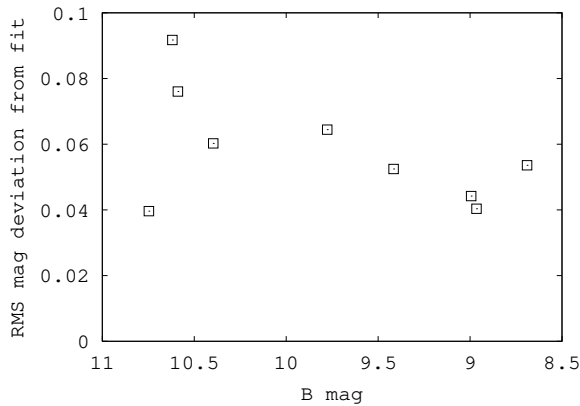,width=8cm}
\caption{Root--mean--square values, in magnitude, for each field star for the 
difference (transformed$-$B) magnitude, plotted against {\it B} magnitude.  
The mean 
RMS value for the stars brighter than B$=$10 is 0.05~mag, which we take as an 
estimate in deriving the magnitude of CF~Oct (B$\sim$9).}
\label{rms_v_B}
\end{figure}

The root--mean--square scatter in the derived (transformed) {\it B} magnitudes 
for 
the 9 field stars over the 1964--1976 interval are shown in 
Figure~\ref{rms_v_B}. These values range from 0.04 to 0.09, with a mean of 
0.058~mag.  The fainter field stars tend to exhibit larger scatter, although 
this may relate more to limitations of the fitting procedure than to photon 
noise. For stars brighter than 10.0, the standard deviations of the derived 
magnitudes are all less than 0.07~mag, with a mean of 0.05~mag.  As CF~Oct 
has B$\sim$9.1, it suggests a precision of $\sim$0.05~mag may be obtainable 
for this star when using the method presented here.

%
%


\subsection{Obtaining the {\it B} magnitudes of CF~Octantis}
\label{sec:B_for_CF}

With the $\beta_{i}$ coefficients determined for each plate, it is a trivial 
matter to apply equation~\ref{eq:with_colour} to the plate magnitude measured 
for CF~Oct and determine the {\it B} magnitude.  The {\it B}$-${\it V} value 
for CF~Oct must of 
course be known in order to be able to apply the calibration equation.  We 
adopt a {\it B}$-${\it V} of 1.10 (Innis et al., 1997).  It is known that the 
{\it B}$-${\it V} of 
CF~Oct is slightly variable, from $\sim$1.10 to 1.15 (Pollard et al., 1989, 
Innis et al., 1997).  Using a {\it B}$-${\it V} of 1.15 would have the effect 
of 
brightening each derived {\it B} magnitude datum by $\sim$0.015~mag.  As the 
likely 
observational error in deriving the magnitude for CF~Oct is around 0.05~mag, 
the uncertainty that arises from using a representative {\it B}$-${\it V} for 
this star can be neglected.

\begin{figure} 
\psfig{file=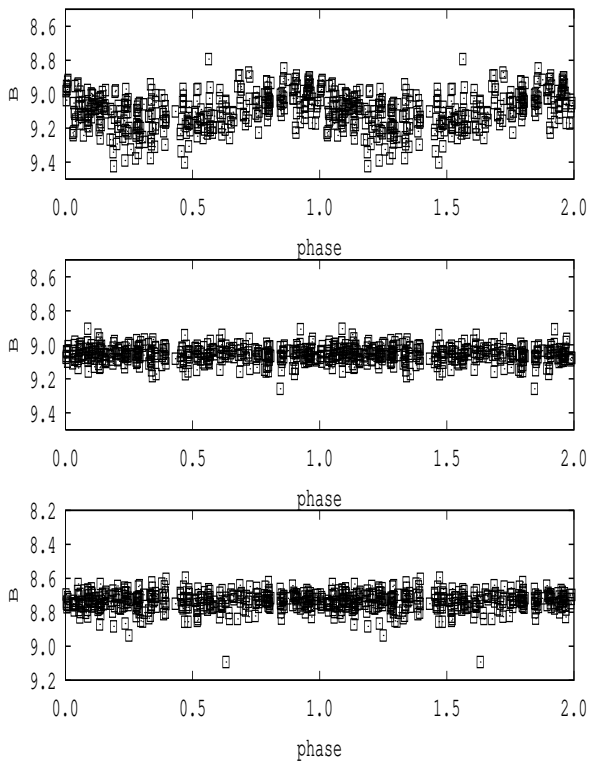,width=8cm,height=10cm}
\caption{Upper panel: Derived {\it B} magnitudes of CF~Octantis, 1964--1976, 
from 
the Bamberg plate archive, using transformation equations of the form of 
equation~\ref{eq:with_colour}. These data are plotted with a period of 
20.15~d, the rotation period of the star determined from a number of years of 
photoelectric photometry in the 1980s. A total of 353 individual measurements 
are presented here. A significant part of the scatter is due to intrinsic 
variation of the star.  Middle and lower panels: {\it B} magnitudes of the field stars 
HD~195460 and HD~196520 respectively, also plotted with the rotation period of CF~Oct.}
\label{all_cf_oneplot}
\end{figure}

Twenty two plate scans proved unsuitable for use, either through plate defects 
(e.g. scratches), poor exposure levels due to cloud, or similar reasons. The 
{\it B} 
magnitudes of CF~Oct derived by use of equation~\ref{eq:with_colour} 
for the remaining 353~plates are plotted in the upper panel of 
Figure~\ref{all_cf_oneplot}, phased with the known 20.15~d period (Pollard et 
al., 1989; Innis et al., 1997).  There is a clear variation with this period. 
The light curve is known to be variable ({\it op cit}), which contributes to a 
significant amount of the scatter in this plot, as we will show.  For 
reference, the middle and lower panels of Figure~\ref{all_cf_oneplot} show the derived 
{\it B} data for HD~195460 (B$-$V=0.20) and HD~196520 (B$-$V=1.05) respectively, also 
plotted with the same period.  The standard deviations of 
the data for these field stars are, to three decimal places, 0.046 and 0.053~mag 
respectively.

\begin{figure*}
\psfig{file=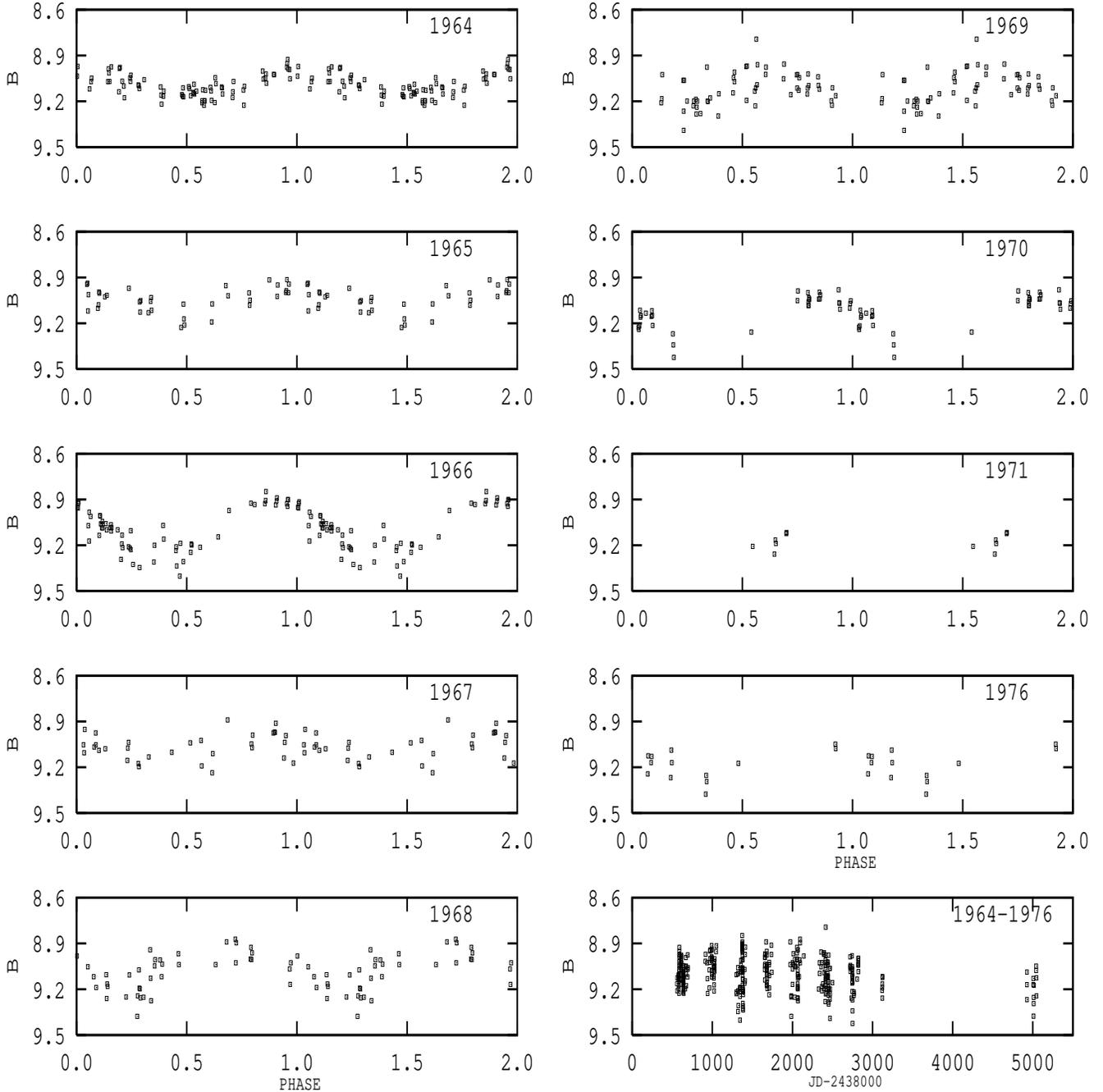,width=18cm,height=18cm}
\caption{Derived {\it B} light curves of CF~Octantis, 1964--1976, from the 
Bamberg 
plate archive, top to bottom, left to right: 1964, 1965, 1966, 1967, 1968, 
1969, 1970, 1971, and 1976. We use a period of 20.15~d for these phase plots 
Lower right panel: Overall {\it B} magnitude variation 
of CF~Oct, plotted against JD$-$2438000.}
\label{separate_phase}
\end{figure*}

The nine individual light curves of CF~Oct from 1964 to 1976 are shown in 
Figure~\ref{separate_phase}, together with a plot showing the overall light 
variation with time.  Good phase cover was obtained in the seasons from 1964 
to 1969 inclusive, with reasonable cover in the other years except for 1971.  
As noted, the light curve of CF~Oct is known to vary, which is presumed to be 
related to changes in starspot properties.  It is however a valid question to 
ask if the data we show in Figure~\ref{separate_phase} represents real changes 
in CF~Oct, or arises simply from some artifact of our analysis, such as an 
incomplete allowance for instrumental effects.  

We cannot, on the basis of these data alone, entirely rule out such effects in 
the CF~Oct data shown in Figure~\ref{separate_phase}.  However, as we have 
ensured, via our transformation procedure, that the derived magnitudes of the 
field stars show a typical variation of $\sim$0.06 mag (one standard deviation 
quoted) during the 1964--1976 span of the data, we have a large measure of 
confidence in internal consistency of the CF~Oct data derived from the Bamberg 
plates.  Of the ten stars measured (nine field stars and CF~Oct), only CF~Oct 
showed any evidence for periodic variation in the $\sim$2 to $\sim$500~d 
range, as tested using the phase dispersion minimisation (PDM) method of 
Stellingwerf (1978); the CF~Oct data showed strong evidence for a $\sim$20~d 
variation (e.g. see Figure~\ref{all_cf_oneplot}), corresponding to the known 
period of variation of the star.  It is clear that the data quality is high 
enough to determine intra--seasonal light curves.  We also believe, based on 
the long--term constancy of the derived magnitudes of the field stars, that 
inter--seasonal changes can also be followed, albeit with an uncertainty of 
perhaps 0.05~mag between seasonal data sets.

\section{Discussion}
\label{discussion}

\subsection{The colour dependence of the photographic plates}

\begin{figure} 
\psfig{file=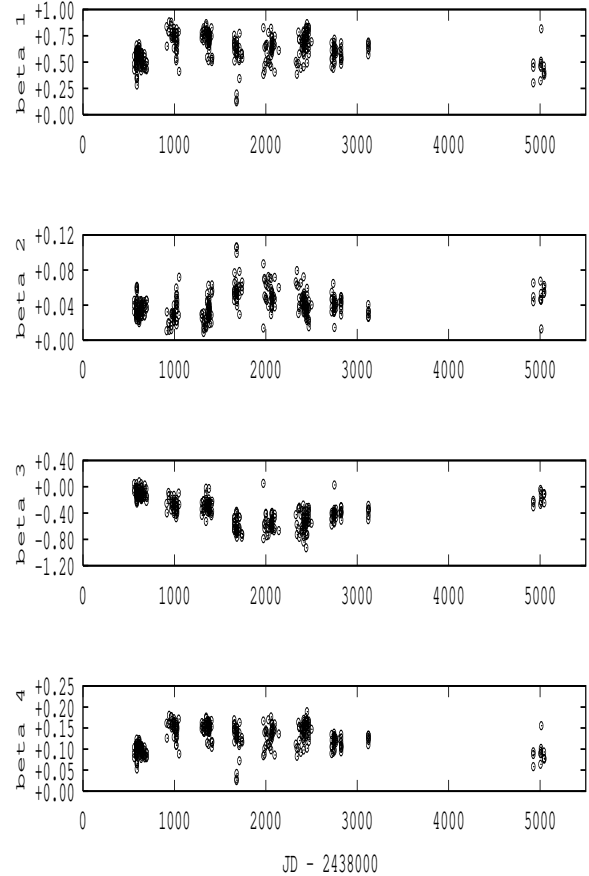,width=8cm,height=12cm}
\caption{Top to bottom: The values of $\beta_{1}$, $\beta_{2}$, $\beta_3$, and 
$\beta_4$, being respectively the coefficients of the M$_{i}$, M$_{i}^{2}$, 
({\it B}$-${\it V}) terms and the additive constant in 
equation~\ref{eq:with_colour}.  See 
the text.}
\label{beta}
\end{figure}

Figure~\ref{beta} shows the values of the $\beta$ parameters 
(equation~\ref{eq:with_colour}) against JD for the Bamberg plates we have 
studied in this work.  From the top panel, it is seen that $\beta_1$ (the term 
linear in instrumental magnitude) is around ten times as big as $\beta_2$ (the 
second-order term). The plate magnitudes are typically around M$_i$$\sim$10, 
hence the contribution to the final magnitude from these two terms is 
approximately equal. The colour term $\beta_3$ shows a systematic 
change over time, with an obvious step between the 1966 South African plates 
and the 1967 New Zealand plates near JD~2439500. The $\beta_3$ values near the 
start and end of the data span are near zero. This appears coincidental 
however, as each season of data is processed independently.  (Indeed, each 
plate is processed independently.)  The $\beta_2$ term shows evidence for changes 
of opposite sense to $\beta_3$, which may suggest these two terms may be interrelated
at some level. However, a varying colour term is present, as evidenced by the raw 
instrumental magnitudes noted earlier.

The variable colour dependency may arise from several factors.  We
understand that plate manufacture and plate development were carefully 
controlled to ensure consistency of spectral response.  It is possible however 
that despite this, some spectral variation may have resulted. Plate storage 
may also be a determining factor.  Equally, ageing of optical coatings and 
surfaces in the cameras may conceivably have an effect, one which could vary 
slowly and systematically over time, as seen in these data.  Atmospheric 
effects may also be significant.  Mt John Observatory in New Zealand, at 
latitude 44$^{\circ}$ S, is nearly 15~degrees south of the latitude of Boyden, 
however the altitudes of the two sites are comparable at $\sim$1000~m. At 
first look, the more southerly location of Mt John may be expected to show a 
reduced colour dependency for the plates for CF~Oct (declination 
$\sim-$80$^{\circ}$), which however is not the case.  Further discussion of 
extinction effects is given in the following section.  For the present we 
conclude that the changes in the colour dependence we have seen are not easily 
explicable.

The data we have presented here come from one 
camera at each site.  As noted, the Bamberg programme used multiple cameras on 
a common mount.  Potentially there is value in comparing plates taken with the 
different cameras at each site. If the colour dependency is mainly due to 
factors associated with the plates, and if these factors applied equally to 
all plates exposed near a particular time, then a similar colour dependence to 
that in Figure~\ref{beta} would be obtained for all cameras.  The same would 
apply if atmospheric effects dominated, although a dependency on airmass (and 
hence declination, as the plates were exposed near upper transit) may be seen. 
Only if the colour dependency were local to each camera (e.g. ageing of 
optical surfaces) would there be different dependencies.

\subsection{Extinction effects}  

The data have been derived without an explicit correction for 
atmospheric extinction.  As all stars are within a 2.75$^{\circ}\times$2.75$^{\circ}$
field, differential extinction effects are smaller than the $\sim$0.05~mag 
plate precision, and can be neglected.  However, colour-dependent 
extinction is also present.  The extinction of a star of 
colour {\it C}, at airmass {\it X}, is approximately:
\begin{equation}
E =  k{'}X - k{''}XC,
\label{approx_extinction}
\end{equation}
where {\it k}${'}$ and {\it k}${''}$ are the first and second order extinction coefficients. More formally, the extinction is (e.g. Young, 1974, Young et al., 1991):
\begin{equation}
\label{young}
E = k{'}X - WRXC - \frac{W(RX)^2}{2},
\end{equation}
where {\it R} is a measure of the atmospheric reddening and {\it W} is proportional to the
square of the optical bandwidth.   The product {\it WR} is equal to {\it k}${''}$.  The third term, 
known as the Forbes effect (e.g. Young, Milone, \& Stagg, 1994) is usually neglected in optical measurements.  For our 
nominal {\it B} band observations, using {\it B}$-${\it V} as the colour 
index, {\it C}, {\it R} 
is the difference in extinction coefficients between the {\it B} and {\it V} 
bands, and 
{\it W} is around 0.28.   When comparing the atmospheric extinction of two 
stars of differing colour, the first term and last terms are maximal when the 
differences in airmass are maximal.  The second term depends on the total 
airmasses, and hence is maximal when the stars are at or near lower transit.  
The Bamberg plates were all obtained within a few hours of upper transit of 
the stars.

As a test to see if unresolved extinction effects were affecting the 
magnitudes we derived, we implemented a scheme based on equation~\ref{young}. 
Because the raw plate magnitudes (or `counts') from the aperture photometry 
have no clear relation to photon counts, we adopted the following method.  The 
extinction for each field star was determined using equation~\ref{young}. We 
adopted extinction coefficients of 0.2 and 0.3 mag~airmass$^{-1}$ for the {\it 
V} 
and {\it B} bands respectively. The catalogued {\it B} magnitude of the star 
was reduced 
by the amount of calculated extinction (i.e. the extra-atmospheric catalogued 
magnitude was corrected to represent the magnitude that would have been 
observed at that time). A preliminary fit between the plate magnitudes and the 
extinction corrected magnitudes (using equation~\ref{eq:with_colour}) was 
performed.  The magnitudes derived from the fit were then corrected to 
extra-atmospheric magnitudes (using the same corrections calculated as above), 
and a second fit between these values and the catalogued {\it B} magnitudes 
was 
found.  

The final magnitudes of the field stars and CF~Octantis were effectively 
identical to those derived without extinction corrections.  This was not 
entirely unexpected.  As noted, differential extinction effects can be 
neglected, as the stars are at small angular separation from each other.  In 
this case, when comparing two stars at similar airmass, only the second term 
in equation~\ref{young} remains: For two stars with a {\it B}$-${\it V} 
difference of 1.0, 
the colour-dependent extinction amounts to about 0.05~mag in B.  This is 
comparable to the precision of measurement of any given star image on a plate. 
Note too it is significantly less than the plate colour dependence term 
$\beta_{3}$ (Figure~\ref{beta}).

We conclude, on this basis, that extinction effects are unlikely to be 
significant, and probably do not account for the appearance of small 
($\sim$0.05~mag) level shifts in the mean levels of the field stars from 
season to season, as seen for example in Figure~\ref{colour}. Additionally, 
the form of equation~\ref{eq:with_colour}, which includes a term in {\it 
B}$-${\it V}, is 
possibly able to incorporate a colour dependent extinction term, so that some 
implicit correction is already applied.  Given the observational error of 
$\sim$0.05~mag from the photographic material, further correction appears 
unecessary.

\subsection{Accuracy and precision of the results}

From the analysis described above, the magnitudes of the field stars and 
CF~Oct can be determined to an estimated 1--standard deviation precision of 
$\sim$0.05~mag from the Bamberg patrol plates.  This is in itself not a new 
result, as it is comparable to the precision reported from iris photometry of 
these plates at the time the patrol was still in operation (e.g. Schoffel, 
1964).  It is clear however that with the advent of fast scanners, and the use 
of very modest computing power, many years of digitised 
plate scans can be processed quickly and easily to recover stellar 
brightnesses of precision comparable to the earlier iris photometer data.

Table~\ref{table:summary} summarises the mean magnitudes and standard 
deviations found for CF~Oct and the 9 field stars chosen for this work for the 
interval 1964--1976, using equation~\ref{eq:with_colour}.  Also included for 
reference are the {\it B} and {\it B}$-${\it V} magnitudes of these stars 
obtained from Tycho 2 
photometry.  The mean values derived using equation~\ref{eq:with_colour} are 
close to the known (`catalogued') {\it B} values.  The mean difference 
(derived$-$catalogued B) is 0.00 mag, with a standard deviation of 0.04~mag.  
The biggest difference is 0.07~mag.  The procedure we use depends on 
accurately known magnitudes, and as noted the derived field star magnitudes 
are not determined independently of the fit.  However, it appears that the 
accuracy with which the field star magnitudes can be recovered is comparable 
to the precision with which individual stars can be measured.

\begin{table}
\caption{Derived {\it B} magnitudes (this study, from 
equation~\ref{eq:with_colour}) found from 353 
Bamberg plates spanning 1964--1976. Also shown are {\it B} and {\it B}$-${\it 
V} magnitudes for 
reference, derived from transforming Tycho 2 photometry.}
\label{table:summary}
\begin{center}
\begin {tabbing}
\
\=Star    \hspace{1.2cm}  \=  \hspace{0.2cm} derived {\it B} \hspace{1.2cm}  
\= \hspace{0.1cm} {\it B}   \hspace{1.3cm}  \=  B-V  \\
\> CF Oct     \>  ~9.10 +/-  0.11 \>  ~(9.1) \> (1.1) \\
\> CP -80 946 \>  10.42 +/-  0.08 \> 10.40  \>  0.23 \\
\> HD 192573  \> 10.54 +/-  0.09 \> 10.59  \>  0.56 \\
\> HD 194005  \> 10.75 +/-  0.07 \> 10.75  \>  1.21 \\
\> HD 195460  \> ~9.06 +/-  0.05 \> ~8.99  \>  0.20 \\
\> HD 196520  \> ~8.73 +/-  0.05 \> ~8.69  \>  1.05 \\
\> CP -80 980 \> 10.64 +/-  0.11 \> 10.62  \>  0.45 \\
\> HD 191289  \> ~8.94 +/-  0.05 \> ~8.97  \>  0.22 \\
\> HD 192950  \> ~9.36 +/-  0.05 \> ~9.42  \>  0.42 \\
\> HD 195291  \> ~9.75 +/-  0.07 \> ~9.78  \>  0.66 \\
\\
\end {tabbing} 
\end{center} 
\end{table}

Clearly, the method we have used is unlikely to be valid unless the field 
stars include a range of {\it B}$-${\it V} (so the colour dependence of the 
plates can be 
determined) and cover a range of {\it B} brighter and fainter than the target 
star.  
Including stars close in brightness to the target star also appears 
desirable, to prevent plate-to-plate variations of a poorly constrained fit 
giving rise to artifacts in the target star's magnitude.

For this initial study we selected and interactively measured only a 
relatively small number of stars on each digitised plate.  There is however no 
reason why automatic aperture photometry could not be performed on all 
suitable star images.  The transformation equations could be derived using the 
majority of the stars, but a small number (say 6 to 10) could be omitted from 
the fit, and their magnitudes derived in the same way as for the target star 
(i.e. as we have done here for CF~Oct).  Their constancy (or otherwise) would 
provide a completely independent check on the quality of the transformation, 
and, by implication, a check on the data quality for the target star.

\subsection{Long--term behaviour of CF Oct}

We intend to present a more detailed analysis of the 1964--1976 CF~Oct data in 
a later paper.  However, we present here a composite photometric history of 
this star, combining these Bamberg data with previously published photometry.  
To do this, we have `converted' the {\it B} data to a proxy for V, by 
subtracting 1.10. This is a representative maximum {\it B}$-${\it V} 
value for CF~Oct (Lloyd Evans \& Koen, 1987; Pollard et al., 1989; Innis et 
al., 1997).  The observed range in {\it B}$-${\it V} of CF~Oct is about 
0.05~mag ({\it op cit.}), which is comparable to the precision of these 
Bamberg data. Also, as noted, we used a constant {\it B}$-${\it V} for CF~Oct 
when deriving the {\it B} magnitudes from the plate magnitudes. For the 
purposes of this comparison using a representative {\it B}$-${\it V} 
value seems justified.

\begin{figure} 
\psfig{file=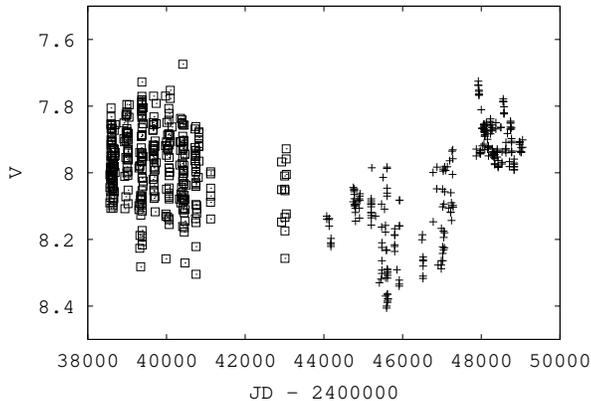,width=8cm,angle=270}
\caption{Long--term photometric history of CF Oct using the {\it B} data 
derived in this work converted to {\it V} assuming a {\it B}$-${\it V} \,for 
CF Oct of 1.10 (shown as squares), combined with previously published 
photoelectric photometry (upright crosses).  See the text.}
\label{cf_history}
\end{figure}

Figure~\ref{cf_history} shows the resulting plot.  The Bamberg proxy {\it V} 
data are shown as squares.  The upright crosses show photoelectric {\it V} 
data from Innis et al. (1983), Lloyd Evans \& Koen (1987), Pollard et al. 
(1989), Innis et al. (1997), and the Hipparcos epoch photometry (ESA, 1997), 
transformed from Hp to V.  The maximum brightness (V$\sim$7.8) and mean level 
of the Hipparcos data is significantly greater than the other photoelectric 
observations.  Tycho photometry (ESA, 1997, not shown) transformed to 
standard {\it V} magnitudes closely matches the Hipparcos data in mean 
brightness levels, and confirms the star really was bright at this time.  It 
is interesting to note that the typical maximum of the proxy {\it V} Bamberg 
data is similar at V$\sim$7.8.  The approximately 9000-d separation between 
these comparable maxima corresponds to $\sim$25~year.  While it is tempting to 
consider the shape of the overall light variation shown in 
Figure~\ref{cf_history} as a stellar activity cycle, clearly a much longer 
time series would be needed to address this issue.  We defer further comment 
to a later paper, except to note that the 9~year cycle suggested by Pollard et 
al. (1989) for CF~Oct is not obvious in these data.

\subsection{Evidence for differential rotation of CF~Oct}

As noted, the rotation period was measured for CF~Oct from a number of 
independent sources from photoelectric photometry between 1979 and 1988, the 
results of which were 20.18~d (Lloyd Evans \& Koen, 1987); 20.15$\pm$0.06~d 
(Pollard et al., 1989); and 20.15~d (Innis et al., 1997).  Innis (1986) 
estimated the period error in the latter data (from 1982 to 1986) as 
$\pm$0.01~d. The phase of minimum light for most of these data (excepting the 
early 1979 South African data) can be kept relatively constant assuming a 
period of 20.15~d.  On this basis, Innis et al. (1997) concluded there was no 
evidence for differential rotation for CF~Oct.

With the increased data now available from the Bamberg archive, it is worth 
briefly revisiting this issue here.  Inspection of 
Figure~\ref{separate_phase}, where the phase plots have been constructed using 
a period of 20.15~d, reveals an apparent smooth migration of minimum light 
from phase $\sim$0.5 in 1964 to $\sim$ 0.2 or 0.3 in 1969.  This suggests
that the rotation period of the dominant spot groups on CF~Oct was slightly 
shorter than 20.15~d.

\begin{figure} 
\psfig{file=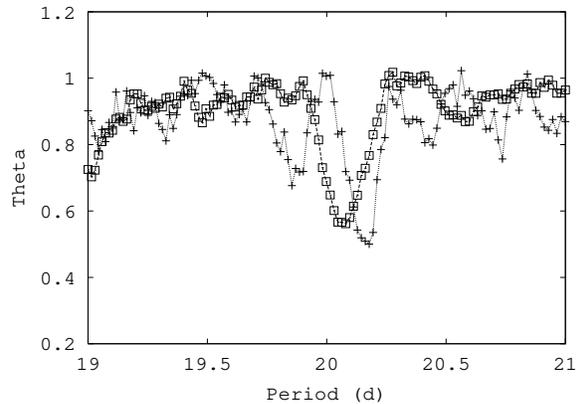,width=8cm,angle=0}
\caption{PDM theta spectra of the 1964--1969 data derived in the current work (squares)
and from photoelectric photometry from 1981--1988 (Lloyd Evans \& Koen, 1987; Pollard et al., 
1988; Innis et al., 1997, upright crosses).}
\label{PDM_comparison}
\end{figure}

We have estimated the rotation period for the interval 1964 to 1969 by forcing 
the phase of minimum light to be constant.  A period near 20.05~d is found, 
but the relatively large observational error per point (compared to 
photoelectric photometry) together with the intrinsic variablity of the light 
curve makes a definitive estimate difficult.  We also performed a PDM 
analysis on the 1964-1976 data, which returned a period near 20.05~d. PDM 
analyses of the 1964-1966 data and, separately, the 1967-1969 data, returned 
periods of 20.07~d and 20.03~d respectively.  Figure~\ref{PDM_comparison} shows as
an illustration of these results the PDM theta spectra for
the 1964--1969 data obtained in this current work (squares), and for the 1981--1988 
photoelectric photometry from Lloyd Evans \& Koen (1987), Pollard et al. (1989) and 
Innis et al. (1997), shown as upright crosses.  The most likely period is indicated by 
the minimum value of theta. The 1964--1969 data show a period near 20.05~d, while the 
photoelectric data from the 1980s return a period near 20.15~d, in agreement with the
results found from choosing periods that minimised the drift in phase of minimum light.

\begin{figure} 
\psfig{file=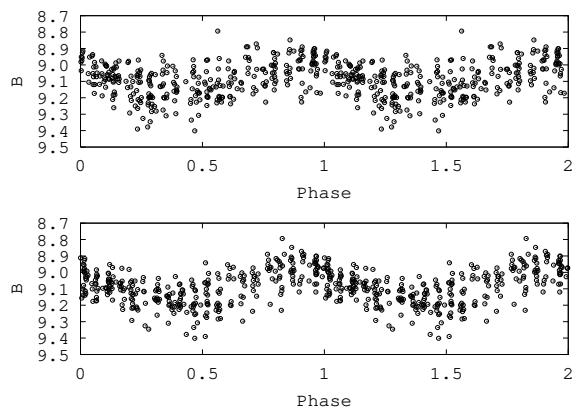,width=8cm,angle=0}
\caption{Light curves of CF Oct using the 1964 to 1969 {\it B} data derived in 
this work for trial periods of 20.15~d (upper panel) and 20.05~d (lower 
panel).}
\label{period_comparison}
\end{figure}

Figure~\ref{period_comparison} shows CF~Oct {\it B} data for 1964 to 1969 
inclusive, 
obtained in this work, plotted with the 20.15~d rotation period (upper panel) 
derived from 
the 1980s era photometry and (lower panel) the same data with a trial period 
of 20.05~d.  There is a slight reduction in scatter in the lower panel 
compared to the upper, largely due to the removal of the migration in phase of 
minimum light noted earlier.  On the basis of the analysis here, and given the 
detection of differential rotation in other active stars, we tentatively 
conclude that the characteristic rotation period of CF~Oct exhibited a small 
but real change between the 1960s and the 1980s.  Further discussion of this 
will be given in a later paper.

\subsection{Comparison of photometry from scanned and digital camera images of 
the plates}

\begin{figure} 
\psfig{file=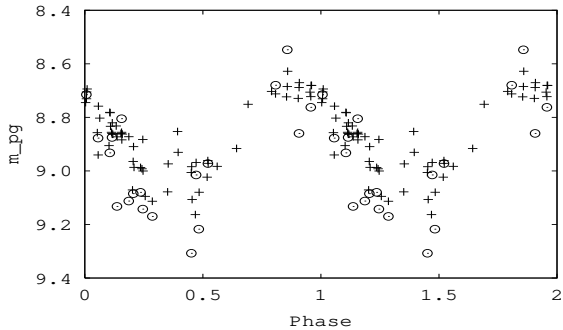,width=8cm,height=4.5cm}
\caption{Phase plot (using a 20.15~d period) showing a comparison of {\it 
photographic} magnitudes derived from the scanned 1966 plates (upright 
crosses), and the results from digital camera images of a subset of the 1966 
plates (open circles) analysed in the preliminary study by Innis et al. 
(2004).}
\label{cam_scan}
\end{figure}

In the introduction we mentioned that preliminary analysis of a subset of the 
1966 plates was carried out by Innis et al. (2004). In that study, commenced 
prior to the installation of the Epson scanner at Bamberg, digital camera 
images of back--illuminated plates were used.  Photographic magnitudes were 
derived for CF~Oct by performing aperture photometry, expressing the results 
as differential magnitudes with respect to HD~196520.  It was known at that 
time that as CF~Oct and HD~196520 are very close in magnitude, colour, and 
angular separation on the sky, any effects due to non-linearity, colour 
dependence, or vignetting over the camera field of view would be minimised.  
What was not appreciated in this early study was just how large the colour 
dependence of the plates would be. Hence it is worth comparing the results 
from the scanned plates, treated in the manner outlined in this paper, with 
the earlier results presented in Innis et al. (2004).  Figure~\ref{cam_scan} 
shows a phase plot of photographic magnitudes for CF~Oct in 1966 derived from 
scanned plate images (upright crosses), together with the results from the 
digital camera study (open circles) of the subset of the 1966 data used in 
Innis et al. (2004).  These latter data were presented as differential 
magnitudes in Innis et al. (2004).  We have added 8.52, the catalogued 
photographic magnitude for HD~196520, to produce this plot.  We use the same 
period and epoch as for Figure~\ref{separate_phase}.

The agreement between the two datasets is reasonably good, as the shape, 
range, and mean brightness are similar.  There is an 
indication of small differences, such as near phase $\sim$0.2 where the 
digital camera data of Innis et al. (2004, shown as circles) appear slightly 
fainter than the scanned data.  Innis et al. (2004) found the digital camera 
images gave measured magnitudes, for stars in the range m$_{pg}$ from 8.5 to 
10.5, that could be directly transformed to catalogued photographic 
magnitudes simply by choosing an appropriate additive constant.  The data 
obtained in the current work (upright crosses) were obtained by transforming 
plate magnitudes to m$_{pg}$ using the methods described in 
sections~\ref{sec:fitting} and \ref{sec:B_for_CF}. Some small differences 
between the datasets almost certainly arise in consequence of this.
 
The digital camera data were obtained with an 8--bit, non--scientific grade 
camera. At least four independent exposures of each plate were made and 
measured, and the results averaged to produce the light curve shown.  The data 
quality from the 14--bit scanner appears to be superior, and we would 
recommend this as the preferable means of plate digitisation if available.  
Where a scanner is not available, digital camera images may still yield 
worthwhile results if care is taken in the analysis.

\subsection{Applicability to other objects}

In principle, photometric histories of a range of other objects could be 
obtained from digitised images of the Bamberg plates, analysed in a manner 
similar to that presented here.  Our study has extended only over the 
approximate magnitude range from {\it B}$\sim$8.7 to 10.8.  It would be 
expected 
that the analysis could be extended to fainter stars, and perhaps also to 
brighter ones.  It may be possible to derive a series of satisfactory 
transformations for each plate, each extending only over 2 magnitudes or so, 
but with a combined range reaching down to the approximate plate limit of 
{\it B}$\sim$14.   

If the plate accuracy and precision we have found, of around 0.05~mag, is 
representative of the entire plate archive, any star with a variation of 
$\sim$0.15~mag or more is in principle capable of yielding a 3--sigma 
detection.  For strictly periodic variables, with a non-varying period, this 
amplitude limit may be lowered if enough data can be obtained.  

The plate exposures are typically of 1~h in duration.  For very rapidly 
varying stars, such as a short-period deeply eclipsing binary, the long 
exposure would tend to average out the variation.  Additionally, unless the 
star is at a high southerly declination (such as CF~Oct) where the camera 
fields have good overlap, the star may only appear on one exposure per night. 

{\it B}$-${\it V} for the star must be known, due to the large colour 
dependency 
of the plate response. Objects that undergo large colour changes would not be 
suitable targets for our approach, unless {\it B}$-${\it V} changes in a 
repeatable, 
phase--locked manner, and can be in some way calculated and corrected for.

Given these caveats, a large number of potential targets remain.  We hope that 
the data we have presented here for CF~Oct will aid in understanding not only 
the behaviour of this particular object and others of its class, but will give 
a hint of the scientific data yet to be found in the world's astronomical 
plate archives.

\section{Conclusions} 

Our analysis of digitised images of the Bamberg Sky Patrol plates of the 
fields of the active K star CF~Oct has shown that estimates of the {\it B} 
magnitude 
of the star, to a precision of $\sim$0.05~mag, are readily obtainable.  We 
have derived a series of light curves from 1964 to 1969 inclusive, with 
partial light curves for 1970 and 1976, and a few points from 1971. These data 
significantly add to the known photometric history of the star. A more 
complete analysis of these data will be presented later.  The results from 
this work indicate that detailed studies of other objects using the Bamberg 
Observatory and other plate archives would be very worthwhile.

\section{Acknowledgments} 
M. Tsvetkov and A. Borisova were supported by the Alexander von Humboldt 
Foundation under the `Pact of stability of South-East Europe' programme and 
grants from BAS/DFG 436-BUL110/120/0-2 and the Bulgarian National Science Fund 
(NFS I-1103/2001).  We thank the Bamberg Observatory staff for hospitality and 
making the archive available. We acknowledge the work of the Bamberg staff who 
collected the data, and archived it for later use.  We used IRAF, from the US 
National Optical Astronomical Observatories, and OCTAVE, by J.W Eaton and 
colleagues, for the data analysis. This research has made use of the on--line 
SIMBAD data facility of the Stellar Data Centre (CDS), Strasbourg, the NASA 
ADS database, and the Sofia Wide Field Plate Database (WFPDB).  D. Coates 
thanks Prof B. Muddle for access to the facilities of the School of Physics 
and Materials Engineering.

\label{lastpage}


\begin{thebibliography}{99} 
 
\bibitem{b1}  Allen C. W.  1973,  Astrophysical Quantities, 3rd ed., 
Athlone Press, London
\bibitem{bessell} Bessell M., 2000, PASP, 112, 961
\bibitem{bondar} Bondar N.I., 1995, AA Suppl. Ser., 111, 259
\bibitem{burke} Burkholder V. 1995, JAAVSO, 23, 127
\bibitem{b13} Collier A.C., 1982, PhD thesis, University of Canterbury, 
New Zealand
\bibitem{ESA} ESA 1997, The Hipparcos and Tycho catalogues, ESA SP-1200
\bibitem{16} Hartmann L., Bopp B.W., Dussault M., Noah P.V., \& Klimke 
A., 1981, Ap.J., 249, 662
\bibitem{b34} Hearnshaw J.B., 1979, in Bateson F.M., Smak J., \&  Urch
I.H., eds, Proq. IAU Colloq. No. 46, Changing Trends in Variable Star 
Research, Univ. Waikato, New Zealand, p.371
\bibitem{b37} Houk N. \& Cowley A.P., 1975.  Michigan Catalogue of 
Two-Dimensional Spectral Types for the HD Stars, Vol.1. (University of 
Michigan, Ann Arbor)
\bibitem{b39} Innis J.L., Coates D.W., Dieters S.W.B., Moon T.T., \& 
Thompson K., 1983, IBVS, No. 2386
\bibitem{b40} Innis J.L., Coates D.W., \& Thompson K., 1997, MNRAS, 289, 515
\bibitem{b41} Innis J.L., Heil P., Thompson K., \& Coates D.W., 2004, 
Publ. ast. Soc. Aus., 21, 284
\bibitem{kroll93} Kroll P. \& Neugebauer P., 1993, AA, 273, 341
\bibitem{b43} Kroll P., la Dous C., \& Brauer H.--J., (eds.), 1999, 
Treasure--Hunting in Astronomical Plate Archives (Frankfurt am Main: Deutsch)
\bibitem{b46} Lloyd Evans T., \& Koen M.C.J., 1987, SAAO Circ., 11, 21
\bibitem{b47} Phillips M.J., \& Hartmann L., 1978, Ap.J. 224, 182
\bibitem{b51} Pollard K.R., Hearnshaw J.B., Gilmore A.C., \& Kilmartin 
P.M., 1989,  JAp\&A, 10, 139
\bibitem{b50} Press W.H., Teukolsky S.A., Vetterling W.T., \& Flannery 
B.P., 1992, Numerical Recipes in Fortran, Cambridge University Press
\bibitem{rucinski} Rucinski S., 1999, in Hearnshaw J.B., Scarfe C.D., eds, IAU 
Colloq. 170, Precise Radial Velocities. Astron. Soc. Pac., San Francisco, p.82
\bibitem{saar} Saar S.H., 1998, IBVS 4580
\bibitem{schoffel} Schoffel E., 1964, IBVS 71
\bibitem{SK} Schoeffel E., \& Koehler U. 1965, IBVS 100
\bibitem{b52} Slee, O.B. Nelson G.J., Stewart R.T., Wright A.E., Innis 
J.L., Ryan S.G., \& Vaughan A.E., 1987a, MNRAS, 229, 659
\bibitem{b53} Slee O.B., Nelson G.J., Stewart R.T., Wright A.E., Jauncey 
D.L., Vaughan A.E., Large M.I., Bunton J.D., Peters W.L., \& Ryan S.G., 
1987b, PASA, 7, 55
\bibitem{stellingwerf} Stellingwerf R.F. 1978, ApJ, 221, 661
\bibitem{b56} Strohmeier W., 1967, IBVS, No.178
\bibitem{stromau} Strohmeier W. \& Mauder H. 1969, Sky. Tel., 37, 10
\bibitem{b57} Vogt N., \& Kroll P., 1999, in Kroll P., la Dous C., \&
 Brauer H.--J., eds., Treasure--Hunting in Astronomical Plate 
Archives, Deutsch, Frankfurt am Main, p.210
\bibitem{young} Young A.T., 1974, in Carleton N., ed, Methods of Experimental 
Physics, Vol 12, Astrophysics; Part A: Optical and Infrared, Academic, New 
York, p.185
\bibitem{young91} Young A.T., Genet R.M., Boyd L.J. et al. 1991, PASP, 103, 221
\bibitem{young94} Young A.T., Milone E.F. \& Stagg C.R. 1994, AA Suppl. Ser., 105, 259

\end{thebibliography}
\end{document}